\documentclass[12pt]{iopart}
\usepackage{epsf}

\def \mref#1{(\ref{#1})}

\begin{document}

\centerline{Quantum Opt. {\bf 2} (1990) 253-265.}

\title[{\rm Quantum Opt. {\bf 2} (1990) 253-265}]
{Generation of discrete superpositions of coherent states in
the anharmonic oscillator model}

\author{A Miranowicz, R Tana\'s and S Kielich}

\address{Nonlinear Optics Division, Institute of Physics, Adam
Mickiewicz University, Grunwaldzka 6, 60-780 Pozna\'n,
Poland}

\date{Received 20 September 1989, in final form 23 February
1990}

\begin{abstract}
The problem of generating discrete superpositions of coherent
states in the process of light propagation through a
nonlinear Kerr medium, which is modelled by the anharmonic
oscillator, is discussed. It is shown that under an
appropriate choice of the length (time) of the medium the
superpositions with both even and odd numbers of coherent
states can appear. Analytical formulae for such
superpositions with a few components are given explicitly.
General rules governing the process of generating discrete
superpositions of coherent states are also given. The maximum
number of well distinguished states that can be obtained for
a given number of initial photons is estimated. The
quasiprobability distribution $Q(\alpha,\alpha^{*},t)$
representing the superposition states is illustrated
graphically, showing regular structures when the component
states are well separated.
\end{abstract}

\maketitle

\section{Introduction}

The generalised coherent states introduced by Titulaer and
Glauber~\cite{1} and discussed by Stoler~\cite{2}, differ
from coherent states by the extra phase factors appearing in
the decomposition of such states into a superposition of Fock
states. Bia\l{}ynicka-Birula~\cite{3} has shown that, under
appropriate periodic conditions, generalised coherent states
go over into discrete superpositions of coherent states. She
also has shown how to calculate the coefficients of such a
superposition. Recently, Yurke and Stoler~\cite{4}, and
Tombesi and Mecozzi~\cite{5} have discussed the possibility
of generating quantum mechanical superpositions of
macroscopically-distinguishable states in the course of the
evolution of the anharmonic oscillator. The anharmonic
oscillator model was earlier used by Tana\'s~\cite{6} to show
a high degree of squeezing for large numbers of photons. The
two-mode version of the model was used by Tana\'s and
Kielich~\cite{7} to describe nonlinear propagation of light
in a Kerr medium, predicting a high degree of what was called
`self-squeezing' of strong light. The comparison of quantum
and classical Liouville dynamics of the anharmonic oscillator
was made by Milburn~\cite{8}, and Milburn and
Holmes~\cite{9}. Kitagawa and Yamamoto~\cite{10} have used
the model in their discussion of the number phase minimum
uncertainty state that can be obtained in a nonlinear
Mach-Zehnder interferometer with a Kerr medium. They
introduced the name `crescent squeezing' for the squeezing
obtained in the model, to distinguish it from `elliptic
squeezing' of an `ordinary' squeezed state. The terms
`crescent' and `elliptic' stem from the shapes of the
corresponding contours of the quasiprobability distribution
$Q(\alpha,\alpha^{*},t)$. The anharmonic oscillator model has
also been discussed by Pe\v{r}inov\'a and Luk\v{s}~\cite{11}
from the point of view of photon statistics and squeezing.
Quantum field superpositions have recently been discussed by
Kennedy and Drummond~\cite{12}, and by Sanders~\cite{13}.
Some properties of generalised coherent states have been
discussed by Vourdas and Bishop~\cite{14}.

In this paper explicit analytical expressions describing
superpositions of up to four coherent states are obtained for
the anharmonic oscillator model with two different orderings
of operators in the interaction Hamiltonian. The maximum
number of clearly distinguishable coherent states in a
superposition is estimated, and the rules describing the
sequence of particular superposition states as time elapses
are given. The results are illustrated graphically for the
coherent initial states with the mean number of photons equal
to 4 or 16, for which the evolution of the quasiprobability
distribution (QPD) $Q(\alpha,\alpha^{*},t)$ is used to
visualise the formation of superposition states.

\section{The anharmonic oscillator model and its evolution}

The anharmonic oscillator model that we discuss in this
paper, is defined by the Hamiltonian
\begin{equation}
\hat H=\hbar\omega \hat a^{\dagger}\hat a+\hat
H_{C}\quad(C=N,S)\label{eq:1}
\end{equation}
where $\hat a$ $(\hat a^{\dagger})$ is the annihilation
(creation) operator and $\hat H_{C}$ describes the chosen
versions of the nonlinear interaction Hamiltonian, which are:
\numparts
\begin{equation}
\hat H_{N}=\frac{1}{2}\hbar\kappa(\hat a^{\dagger})^{2}\hat
a^{2} =\frac{1}{2}\hbar\kappa\hat{n}(\hat{n}-1)\label{eq:2a}
\end{equation}
 \begin{equation}
\hat H_{S}=\frac{1}{2}\hbar\kappa(\hat a^{\dagger}\hat a)^{2}
=\frac{1}{2}\hbar\kappa\hat{n}^{2}.\label{eq:2b}
\end{equation}
\endnumparts
Here $\hat{n}=\hat a^{\dagger}\hat a$ is the number of
photons operator and $\kappa$ is the coupling constant, which
is real and can be related to the nonlinear susceptibility
$\chi^{(3)}$ of the medium if the anharmonic oscillator is
used to describe the propagation of laser light in a
nonlinear Kerr medium. Both versions of the interaction
Hamiltonian are in use, depending on the authors. The
difference between them seems to be trivial because it means
a change in the free oscillator frequency of the oscillator.
When the homodyne detection of squeezing is to be applied,
however, this extra phase shift can be significant in the
long-time limit~\cite{15}. Thus, the question arises: which
version is to be used in a particular physical situation?
From the point of view of quantum-classical correspondence
the normal ordering is preferable because it makes the
transition from the quantum to the classical description via
coherent states quite transparent and preserves the classical
meaning of the nonlinear susceptibility of the medium.
However, since both versions are used in the literature, in
this paper we consider both of them separately in order to
make the difference more explicit.

Since the number of photons $\hat{n}$ is a constant of motion
(it commutes with both versions of the interaction
Hamiltonian) the state evolution of the system is described,
in the interaction picture, by the Schr\"odinger equation
\begin{equation}
i\hbar\frac{\rm d}{{ d}t}\hat U_{C}(t)=\hat H_{C}\hat
U_{C}(t)\label{eq:3}
\end{equation}
where $\hat U_{C}(t)$ is the time evolution operator. In the
propagation problem of light propagating in a Kerr medium,
one can make the replacement $t=-z/v$ to describe the spatial
evolution of the field, instead of the time evolution. The
solution of equation~\mref{eq:3} is given by
\begin{equation}
\hat U_{C}(z)=\exp\left(\frac{iz}{\hbar\upsilon}\hat
H_{C}\right)
=\exp\left(i\hat{\theta}_{C}(\hat{n})\right)\label{eq:4}
\end{equation}
where\[
\hat{\theta}_{N}(\hat{n})=\frac{\tau}{2}\hat{n}(\hat{n}-1)\]
or
\begin{equation}
\hat{\theta}_{S}(\hat{n})=\frac{\tau}{2}\hat{n}^{2}\label{eq:5}
\end{equation}
 and
\begin{equation}
\tau=\kappa z/\upsilon\label{eq:6}
\end{equation}
is a dimensionless length of the medium (or time in the time
domain). Since the difference between the time and the
spatial descriptions is trivial (the main effect is the
change in sign), we write formulae for the spatial
description but use the terms time or length interchangeably
in the text.

If the state of the incoming beam is a coherent state
$|\alpha_{0}\rangle$, the resulting state of the outgoing
beam is given by
\begin{equation}
|\psi(\tau)\rangle=\hat
U_{C}(\tau)|\alpha_{0}\rangle.\label{eq:7}
\end{equation}
Using the well known decomposition of the coherent state
$|\alpha_{0}\rangle$,
\begin{equation}
|\alpha_{0}\rangle=\exp\left(-\frac{|\alpha_{0}|^{2}}{2}\right)
\sum_{n=0}^{\infty}\frac{\alpha_{0}^{n}}{\sqrt{n!}}|
n\rangle\label{eq:8}
\end{equation}
we obtain from equations~\mref{eq:4} and~\mref{eq:7}:
\begin{equation}
|\psi(\tau)\rangle=|\alpha_{0},\tau\rangle
=\exp\left(-\frac{|\alpha_{0}|^{2}}{2}\right)\sum_{n=0}^{\infty}\frac{\alpha_{0}^{n}}{\sqrt{n!}}\exp\left[i\theta_{C}(n)\right]|
n\rangle\label{eq:9}
\end{equation}
where
\begin{equation}
\theta_{N}(n)=\frac{\tau}{2}n(n-1),\quad\theta_{S}(n)
=\frac{\tau}{2}n^{2}.\label{eq:10}
\end{equation}
Because of the presence of the additional phases
$\theta_{N}(n)$ or $\theta_{S}(n)$, the resulting state is a
generalised coherent state~\cite{1,2} which can be, under
certain conditions~\cite{3}, a discrete superposition of
coherent states. Some of these superpositions will be given
in the next section.

A good representation of the field state resulting during the
evolution of the anharmonic oscillator is the
quasiprobability distribution $Q(\alpha,\alpha^{*},t)$
defined as, see~\cite{8}:
\begin{equation}
Q(\alpha,\alpha^{*},t)
={\rm Tr}(\hat{\rho}(\tau)|\alpha\rangle\langle\alpha|)
=\langle\alpha|\hat{\rho}(\tau)|\alpha\rangle.\label{eq:11}
\end{equation}
 This function satisfies the relations
\begin{equation}
\int Q(\alpha,\alpha^{*},t)\frac{{\rm d}^{\,2}\alpha}{\pi}
=1\label{eq:12}
\end{equation}
and
\begin{equation}
0\leq Q(\alpha^{*},\alpha,t)\leq 1.\label{eq:13}
\end{equation}
The properties of this function for the anharmonic oscillator
both in the classical and the quantum descriptions of the
oscillator have been discussed by Milburn~\cite{8} and
Milburn and Holmes~\cite{9}.

In the case of the initial state $|\alpha_{0}\rangle$, the
$Q$-function has the form
\begin{equation}
Q(\alpha,\alpha^{*},0)=\exp(-|\alpha-\alpha_{0}|^{2})\label{eq:14}
\end{equation}
which is a Gaussian bell centred on $\alpha_{0}$.

Since the state of the outgoing field is given by
equation~\mref{eq:9}, its density operator is
$\hat{\rho}=|\psi(\tau)\rangle\langle\psi(\tau)|$, and the
corresponding quasiprobability distribution is given
by~\cite{8,10}:
\begin{eqnarray}
Q_{C}(\alpha,\alpha^{*},\tau) &
=&\langle\alpha|\psi(\tau)\rangle\langle\psi(\tau)
|\alpha\rangle\nonumber\\
 & =&\exp(-|\alpha|^{2}-|\alpha_{0}|^{2})\Big|
 \sum_{n=0}^{\infty}\frac{(\alpha^{*}\alpha_{0})^{n}}{n!}
 \exp\left[i\theta_{C}(n)
 \right]\Big|^{2} \label{eq:15}
\end{eqnarray}
where $\theta_{C}$ are given by equations~\mref{eq:10}. This
quasiprobability distribution will be illustrated graphically
for some specific values of $\tau$ to show the formation of
the superposition states.

According to equations~\mref{eq:9} and~\mref{eq:15}, it is
clear that both the state itself and the quasiprobability
distribution exhibit periodic behaviour, however, the two
versions of the anharmonic oscillator have different periods.
Since $n(n-1)$ is always an even number (contrary to $n^{2}$,
which can be odd) we have from~\mref{eq:10} that the period
for the normally ordered version of the interaction
Hamiltonian is one half of the period for the `squared'
version. We have
\begin{equation}
Q_{C}(\alpha,\alpha^{*},\tau+T)=Q_{C}(\alpha,\alpha^{*},\tau)
\label{eq:16}
\end{equation}
with the periods
\begin{equation}
T_{N}=2\pi,\qquad T_{S}=4\pi\label{eq:17}
\end{equation}

The periodic behaviour of the quasiprobability
distribution~\mref{eq:15}, or the state~\mref{eq:9}, can be
observed in the long-time (or long-length) limit. Estimates
based on realistic values of the non-linear susceptibility of
the medium give, for a length of the medium of the order of
one metre, values of $\tau$ of the order of
$10^{-6}$~\cite{7}. This makes it rather unrealistic to
observe periodic behaviour, at least in the case of the Kerr
medium. Such a periodic behaviour is, on the other hand, an
essential feature of the quantum dynamics of the system and,
thus, worth studying in its own right. Some quantum features
of the system such as squeezing are more likely to be
observed for a large number of photons in the short-time
limit~\cite{6,7}. The generation of superpositions of the
macroscopically-distinguishable states, however, needs rather
long evolution times.

\section{Generation of discrete superpositions of coherent
states}

The state of the field obtained as a result of the evolution
of the anharmonic oscillator, which is given by
equation~\mref{eq:9}, is a generalised coherent
state~\cite{1,2}. When the phases $\theta_{C}(n)$ satisfy
periodic conditions
\begin{equation}
\exp\left[i\theta_{C}(n+N)\right]=\exp\left[i\theta_{C}(n)
\right]\label{eq:18}
\end{equation}
for every $n$, with $N$ being an arbitrary positive integer
number, the state~\mref{eq:9} can be represented as a
discrete superposition of $N$ coherent states~\cite{3}
\begin{equation}
|\psi(\tau)\rangle=|\alpha_{0},\tau\rangle
=\sum_{k=1}^{N}c_{k}|\exp(i\varphi_{k})\alpha_{0}\rangle\label{eq:19}
\end{equation}
where the phases $\varphi_{k}$ and the coefficients $c_{k}$ are
to be found. For a specific choice of the time
(length)\begin{equation} \tau_{S}=T_{S}/N=4\pi/N\label{eq:20}
\end{equation}
the periodic conditions~\mref{eq:18} are satisfied and, for
$N$ odd, the superposition~\mref{eq:19} can be found directly
according to the formulae given by
\begin{equation} \varphi_{k}=2\pi
k/N\quad\quad k=1,2,...,N\quad\quad N\; {\rm odd},\label{eq:21}
\end{equation}
and the coefficients $c_{k}$ can be found from the following
system of $N$ equations:
\begin{equation}
\hspace{-1.5cm}\sum_{k=1}^{N}c_{k}\exp(in\varphi_{k})=
\exp\left[i\theta_{S}(n)\right]=\exp\left[i(2\pi/N)n^{2}\right]
\quad\quad n=0,1,...,N-1.\label{eq:22}
\end{equation}
With the choice~\mref{eq:20} of the time (length), there is
some additional symmetry that allows us to reduce the number
of equations that are needed for finding the coefficients
$c_{k}$. Using the relations
\begin{equation}
\exp\left[i\theta_{S}(N-n)\right]=\exp\left[i\theta_{S}(n)
\right]\label{eq:23}
\end{equation}
one easily finds that
\begin{equation}
c_{N-k}=c_{k}\label{eq:24}
\end{equation}
which means that the number of equations is reduced to
$\frac{1}{2}(N-1)+1$. In the case of odd $N$, we obtain a
superposition of $N$ coherent states with their $\alpha$
satisfying the relation $|\alpha_{k}|=|\alpha_{0}|$.

When $N$ is even, the following relations hold:
\begin{equation}
\exp\left[i\theta_{S}(n+N/2)\right]=(-1)^{N/2}
\exp\left[i\theta_{S}(n)\right]\label{eq:25}
\end{equation}
 \begin{equation}
\exp\left[i\theta_{S}(N/2-n)\right]=(-1)^{N/2}
\exp\left[i\theta_{S}(n)\right].\label{eq:26}
\end{equation}

If the relation~\mref{eq:25} is applied to
equation~\mref{eq:22}, it becomes evident that, depending on
whether $N/2$ is odd or even, only the coefficients with odd
or even $k$ survive. This reduces the number of equations by
one half. Instead of $N$ equations that are needed in the
general case, there are only $N/2$ equations and,
correspondingly, the resulting superposition has only
$N/2$ coherent states. In effect we obtain:

(i) For $N$ even and $N/2$ odd
\begin{equation}
\varphi_{k}=2\pi(2k-1)/N\quad\quad k=1,2,...,N/2\label{eq:27}
\end{equation}
\begin{equation}
\hspace{-1.5cm}
\sum_{k=1}^{N/2}c_{2k-1}\exp\left[in\varphi_{k}\right]
=\exp\left[i\theta_{S}(n)\right]
=\exp\left[i(2\pi/N)n^{2}\right]\quad\quad
n=0,1,...,\frac{1}{2}N-1\label{eq:28}
\end{equation}
and, when the relation~\mref{eq:26} is exploited, the
following relation between the coefficients is found:
\begin{equation}
c_{N-(2k-1)}=c_{2k-1}\label{eq:29}
\end{equation}
which reduces the number of equations to
$\frac{1}{2}(\frac{1}{2}N-1)+1.$

(ii) For $N$ even and $N/2$ even
\begin{equation}
\varphi_{k}=2\pi(2k/N)\quad\quad k=1,2,...,N/2\label{eq:30}
\end{equation}
 \begin{equation}
\sum_{k=1}^{N/2}c_{2k}\exp\left[in\varphi_{k}\right]=
\exp\left[i\theta_{S}(n)\right]=\exp\left[i(2\pi/N)n^{2}\right]
\label{eq:31}
\end{equation}
and again the relation~\mref{eq:26} leads to
\begin{equation}
c_{N-2k}=c_{2k}\label{eq:32}
\end{equation}
reducing the number of equations to $\frac{1}{4}N+1$.

Of course, the numbering of the coefficients $c_{2k-1}$ and
$c_{2k}$ can be replaced by $c_{k}$. However, we keep the
above notation for $c_{k}$ in order to indicate their origin
and to show clearly their symmetry.

It is evident from the results obtained above that the
symmetry of the system under consideration plays a crucial
role in reducing the problem of finding the superposition
states. It is also clear from~\mref{eq:21} to~\mref{eq:27}
that superpositions of say three states appear both for $N=3$
and for $N=6$. These are, however, different states.
Comparison of~\mref{eq:21} and~\mref{eq:27} shows that the
phases of the two superpositions differ by $\pi/3$, which
means reflection with respect to the Im$\alpha$ axis. In
fact, if we take the time equal to $2T_{S}/6=T_{S}/3$
in~\mref{eq:28}, we easily recover the state obtained
from~\mref{eq:22} for $T_{S}/3$, if we simultaneously replace
$\varphi_{k}$ by $2\varphi_{k}$. So, the number of components
in the superposition depends on what fraction of the period
we take for the time. If the fraction of the period is $m/N$,
assuming that this fraction cannot be reduced, the number of
components is equal to $N$ for $N$ odd, and $N/2$ for $N$
even. If the fraction $m/N$ can be reduced, the above rules
must be applied to the reduced fraction. It is also not
difficult to prove that coefficients $c_k$ obtained for
$(N-k)/N$ are complex conjugates of those for $k/N$. These
are general rules governing the process of generation of the
discrete superpositions of coherent states during the
evolution of the anharmonic oscillator. Of course, when the
evolution starts at time $\tau=0$, the superpositions with
large numbers of components will appear first, while the
superpositions with few components cannot appear before the
time approaches half the period or a fraction of this with
small denominators (2,3,4,...). Since the superpositions with
a small number of coherent components are most interesting in
the discussion of macroscopically-distinguishable states, we
give here some examples of such states obtained with the use
of the formulae derived in this section.

For $N=2$, according to the rules, there is only one coherent
state in the superposition, which according to~\mref{eq:27}
and~\mref{eq:28} is equal to
\begin{equation}
|\alpha_{0},T_{S}/2\rangle_{S}=|\alpha_{0},2\pi\rangle_{S}
=|\exp(i\pi)\alpha_{0}\rangle=|-\alpha_{0}\rangle.\label{eq:33}
\end{equation}
 When $N=4$, the number of components is equal to two and, according
to~\mref{eq:30} and~\mref{eq:31} we have
\begin{equation}
|\alpha_{0},T_{S}/4\rangle_{S}=|\alpha_{0},\pi\rangle_{S}
=\frac{1}{\sqrt{2}}\left[\exp(-i\pi/4)|-\alpha_{0}\rangle
+\exp(i\pi/4)|\alpha_{0}\rangle\right].\label{eq:34}
\end{equation}
Equations~\mref{eq:33} and~\mref{eq:34} are the results given
by Yurke and Stoler~\cite{4} in their discussion of the
problem of generation of a superposition of macroscopically-distinguishable
states. For $\tau=3T_{S}/4$ we obtain the
state with coefficients that are complex conjugates of the
coefficients in~\mref{eq:34}.

For $N=3$, we have obtained from~\mref{eq:21}
and~\mref{eq:22} the following superposition:
\begin{eqnarray}
|\alpha_{0},T_{S}/3\rangle_{S}
=|\alpha_{0},4\pi/3\rangle_{S}&=&\frac{1}{\sqrt{3}}
\big[\exp(-i\pi/6)|\exp(i2\pi/3)|\alpha_{0}\rangle
\nonumber \\
&& +\exp(-i\pi/6)|\exp(-i2\pi/3)\alpha_{0}
+i|\alpha_{0}\rangle\big]\label{eq:35}
\end{eqnarray}
and for $\tau=2T_{S}/3$ the state with the complex-conjugated
coefficients is obtained.

For $N=6$, according to~\mref{eq:27} and~\mref{eq:28}, we
have
\begin{eqnarray}
\hspace{-1.5cm}
|\alpha_{0},T_{S}/6\rangle_{S}=|\alpha_{0},2\pi/3\rangle_{S}
&=&\frac{1}{\sqrt{3}}
\big[\exp(i\pi/6)|\exp(i\pi/3)|\alpha_{0}\rangle
\nonumber \\
&&+\exp(i\pi/6)|\exp(-i\pi/3)\alpha_{0}\rangle
-i|-\alpha_{0}\rangle\big].\label{eq:36}
\end{eqnarray}
This state is different from~\mref{eq:35}, but the
state~\mref{eq:35} is obtained for $\tau=2T_{S}/6=T_{S}/3$.
For $N=8$, we have obtained:
\begin{eqnarray}
\hspace{-1.5cm}
|\alpha_{0},T_{S}/8\rangle_{S}&=&|\alpha_{0},\pi/2\rangle_{S}
\nonumber \\
&=&\frac{1}{2}\left[|
i\alpha_{0}\rangle-\exp(i\pi/4)|-\alpha_{0}\rangle
+|-i\alpha_{0}\rangle+\exp(i\pi/4)|\alpha_{0}\rangle\right]
\label{eq:37}
\end{eqnarray}
\begin{eqnarray}
\hspace{-1.5cm}
|\alpha_{0},3T_{S}/8\rangle_{S}&=&|\alpha_{0},3\pi/2\rangle_{S}
\nonumber \\
&=&\frac{1}{2}\left[|
i\alpha_{0}\rangle+\exp(-i\pi/4)|-\alpha_{0}\rangle
+|-i\alpha_{0}\rangle-\exp(-i\pi/4)|\alpha_{0}\rangle\right].
\label{eq:38}
\end{eqnarray}
The results~\mref{eq:37} and~\mref{eq:38} are interesting
because they correspond to the plots of the QPD
$Q(\alpha,\alpha^{*},t)$ given, for $\alpha_{0}=2.0$, by
Milburn~\cite{8}, from which four Gaussian peaks of the QPD
are clearly visible. The two-peak structure corresponding to
the state~\mref{eq:35} is also evident. Knowing the
superposition states makes the interpretation of the
multipeak structure of the QPD quite transparent. The
quasiprobability distribution has four peaks because it
represents a superposition state composed of four coherent
states. This, of course, is true when the component states
are well separated and the interference terms are negligible.
Thus, the question arises: when can the components of the
superposition be considered as well separated? To answer this
question we have to remember that the Gaussian
quasiprobability distribution~\mref{eq:14} representing a
coherent state has a finite width. If we assume, somewhat
arbitrarily, that the states are well separated when the
distance between their Gaussian peaks in the complex
$\alpha$-plane is equal to the diameter of the contour
obtained when the section of the Gaussian bell is made at 0.1
of its height, the diameter is then estimated by the value
$2(\ln10)^{1/2}\cong3.03.$ On the other hand, all the
coherent states entering the superposition have their $\alpha_{k}$
parameters such that $|\alpha_{k}|=|\alpha_{0}|.$ This means
that all the Gaussian peaks representing such states are
distributed regularly around a circle of radius
$|\alpha_{0}|$ in the complex $\alpha$-plane. So, the maximum
number $N_{\max}$ of the well-separated Gaussians for given
$|\alpha_{0}|$ can be estimated by
\begin{equation}
N_{\max}\cong2\pi|\alpha_{0}|/[2(\ln10)^{1/2}]\cong
2.07|\alpha_{0}|.\label{eq:39}
\end{equation}
This estimation gives for $|\alpha_{0}|=2$ that the maximum
number of well-separated peaks in the QPD
$Q(\alpha,\alpha^{*},t)$ is four. These are the four peaks
obtained by Milburn~\cite{8}. In fact, the five-peak or even
six-peak structure of the QPD can still be identified when
the proper $\tau$ is taken, but the peaks are not well
separated and their shapes are strongly affected by the
interference terms.

To make this point clearer we write down the analytical
expression for the QPD of the discrete superposition of
coherent states which can be split into the sum of pure
Gaussians and another sum describing the interference terms
\begin{equation}
Q=Q_{\rm Gauss}+Q_{\rm int}\label{eq:40}
\end{equation}
where
\begin{eqnarray}
Q_{\rm Gauss}=\sum_{k=1}^{N}Q_{k}
\nonumber\\ Q_{\rm
int}=\sum_{k>l}2\, {\rm Re}\, Q_{kl}\label{eq:41}
\end{eqnarray}
with
\begin{equation}
Q_{k}=|c_{k}|^{2}\exp\left(-|\alpha-\alpha_{l}|^{2}\right)
\label{eq:42}
\end{equation}
\begin{eqnarray}
2\, {\rm Re}\, Q_{kl} & =&2|c_{k}||c_{l}|
\exp\left(-\frac{1}{2}|\alpha-\alpha_{k}|^{2}
-\frac{1}{2}|\alpha-\alpha_{l}|^{2}\right) \nonumber
\\
&& \times\cos\left[\gamma_{k}-\gamma_{l}
+|\alpha||\alpha_{0}|(\sin\Delta\varphi_{k}-\sin\Delta\varphi_{l})
\right]) \label{eq:43}
\end{eqnarray}
In equation~\mref{eq:43} we have used the notation
\begin{equation}
c_{k}=|c_{k}|\exp(i\gamma_{k}),\quad\alpha
=|\alpha|\exp(i\varphi),\quad\alpha_{0}
=|\alpha_{0}|\exp(i\varphi_{0})\label{eq:44}
\end{equation}
and
\begin{equation}
\Delta\varphi_{k}=\varphi_{k}-\varphi_{0}-\varphi.\label{eq:45}
\end{equation}

\begin{figure}
\epsfxsize=11cm\hspace{2.5cm}\epsfbox{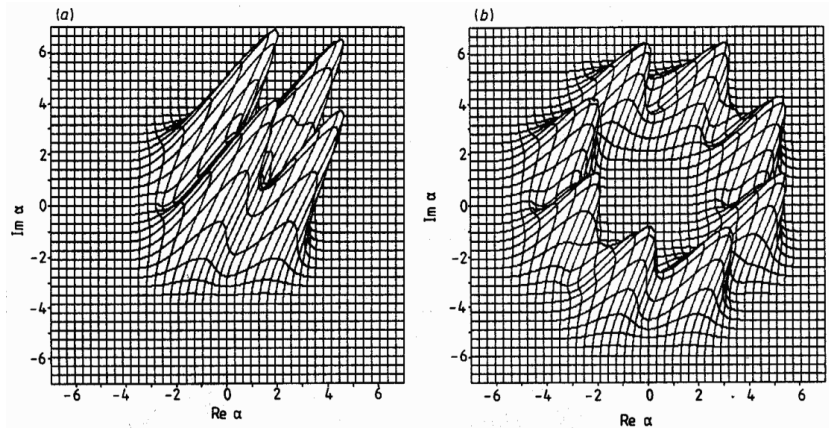}
\caption{Plots of the quasiprobability distributed
$Q(\alpha,\alpha^{*},t)$; (a) for $\alpha_{0}=2.0$ and
$\tau=T_{N}/4$ (the four-peak structure) and (b) for
$\alpha_{0}=4.0$ and $\tau=T_{N}/8$ (the eight-peak
structure).}
\end{figure}

\begin{figure}
\epsfxsize=8cm \centerline{\epsfbox{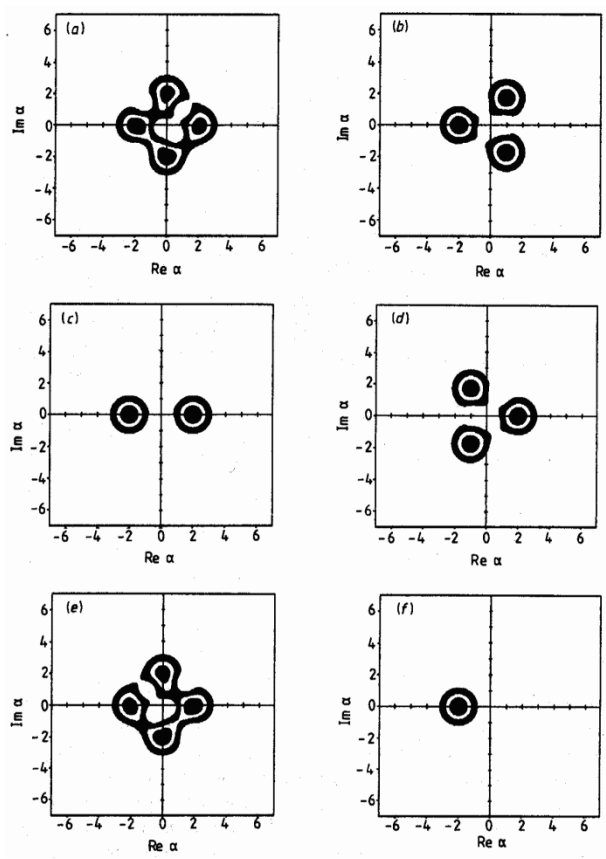}}
\caption{Contours of the quasiprobability distribution for
the `squared' version of the anharmonic oscillator obtained
for sections at $\frac{1}{4},\frac{1}{2}$ and $\frac{3}{4}$
of the height. The parameters are $\alpha_{0}=2.0$, and (a)
$\tau$ equal to $T_{S}/8$, (b) $T_{S}/6$, (c) $T_{S}/4$, (d)
$T_{S}/3$, (e) $3T_{S}/8$ and (f) $T_{S}/2$.}
\end{figure}

\begin{figure}
\epsfxsize=8cm \centerline{\epsfbox{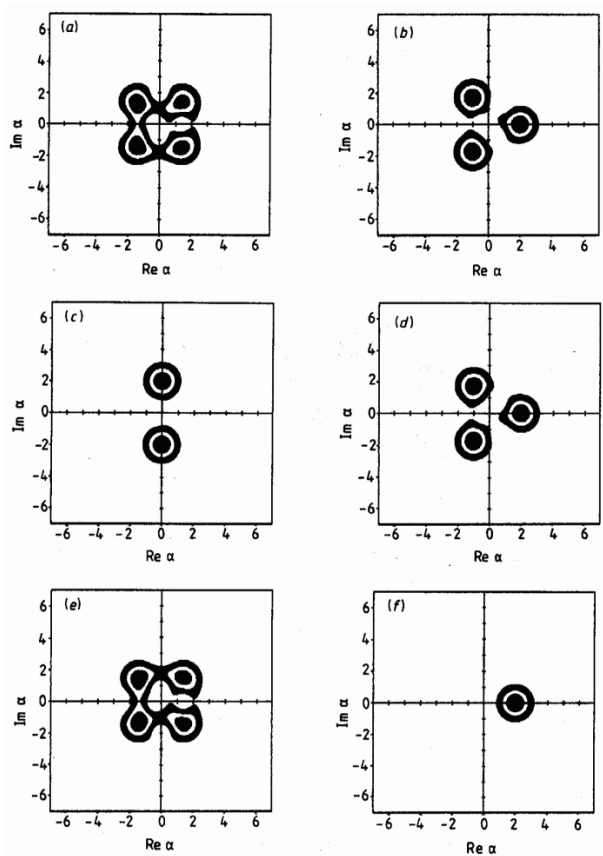}} \caption{The
same as in figure 2 but for the normally ordered version of
the anharmonic oscillator. To visualise the differences
between the two versions of the model, the same
moments,$\tau$, are taken, that is (a) $\tau=T_{S}/8$, (b)
$T_{S}/6$, (c) $T_{S}/4$, (d) $T_{S}/3$, (e) $3T_{S}/8$ and
(f) $T_{S}/2$.}
\end{figure}

\begin{figure}
\epsfxsize=8cm \centerline{\epsfbox{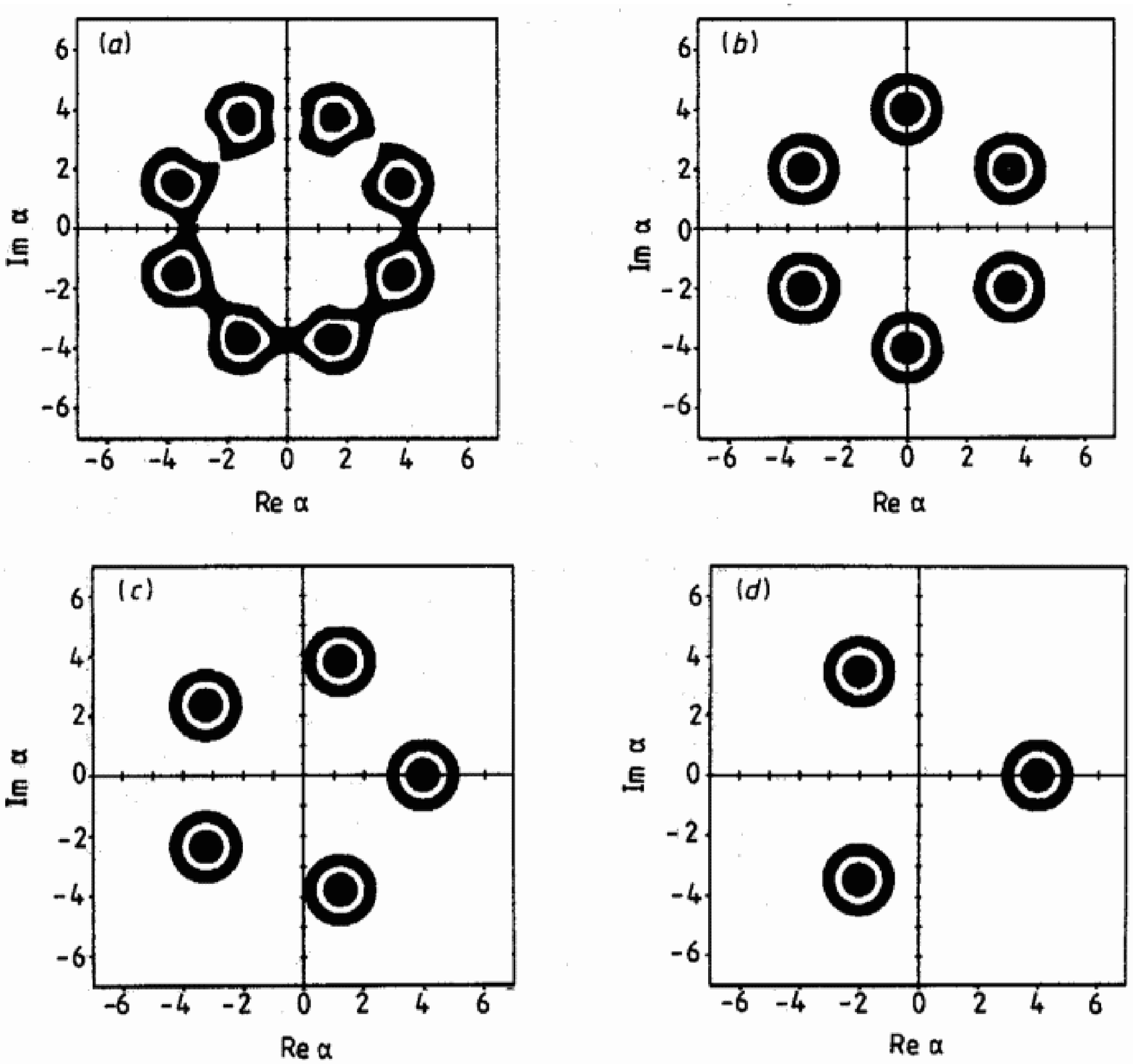}} \caption{The
same sections of the QPD for the normally ordered version of
the anharmonic oscillator as in figure 3, but for the
parameters: $\alpha_{0}=4.0$ and (a) $\tau$ equal to
$T_{N}/8$, (b) $T_{N}/6$, (c) $T_{N}/5$, and (d) $T_{N}/3$.}
\end{figure}

In deriving equations~\mref{eq:40}--\mref{eq:45} the
superposition state~\mref{eq:19} has been used. It is clear
from~\mref{eq:43} that due to oscillations of the cosine
function the interference terms can have a number of peaks.
However, due to the exponential factor the amplitudes of
these peaks are very small whenever the states $k$ and $l$
are well separated. `Well separated' means here that
$|\alpha_{k}-\alpha_{l}|^{2}\gg1$. This gives us another
criterion for good separation of states. In the following we
illustrate the formation of the superposition states by
showing pictures of their QPD for special situations.

Before doing this, however, we have to make some comments on
the behaviour of the normally ordered version of the
anharmonic oscillator. Our analytical
formulae~\mref{eq:33}--\mref{eq:38} are for the `squared'
version of the anharmonic oscillator; however, it is clear
from~\mref{eq:10} that
$\theta_{N}(n)=\theta_{S}(n)-\frac{1}{2}\tau n.$ If this is
inserted into equation~\mref{eq:9}, it is seen that the
superposition states obtained from the normally ordered
version acquire an additional phase $\varphi=-\tau/2.$ Thus,
the only change that is needed to obtain the results for this
version is the replacement of the $\varphi_{k}$ by the
$\varphi-\frac{1}{2}\tau$ in all the formulae obtained in
this section. Geometrically this means the rotation of the
QPD picture by the angle $\varphi=-\tau/2$ in the complex
$\alpha$-plane, without any change in its shape. This rotation may,
nevertheless, change in an essential way the general view of
the QPD. For example, the QPD representing the
state~\mref{eq:33} is a Gaussian centred at $-\alpha_{0}$,
and after the rotation by $\pi$ it becomes a Gaussian centred
at $+\alpha_{0}$, that is at the initial value of
$\alpha_{0}$. There is no single Gaussian peak at
$\mathbf{-\alpha_{0}}$ for the normally ordered version. This
is a general rule, related to the fact that the period for
the normally ordered version is one half of the period for
the `squared' version; this will be convincingly shown in the
figures.

In figure 1 we show two examples of the QPD. The first
example exhibits four Gaussian peaks for the case considered
by Milburn of $\alpha_{0}=2.0$. The second example shows the
eight Gaussian peaks in the case $\alpha_{0}=4.0$, which
according to our estimate~\mref{eq:39} is the maximum number
of well-separated Gaussians in this case. Both examples
represent quite regular shapes confirming the formation of
superposition states with a definite number of components.
All examples presented in our pictures are obtained
numerically from the expression~\mref{eq:15}. In figure 2
contours of the sections at
$\frac{1}{4},\frac{1}{2},\frac{3}{4}$ of the height of the
QPD for the `squared' version of the anharmonic oscillator
are presented, for $\alpha_{0}=2.0$, showing that for the
maximum number of well-separated states $N_{\max}=4$
[estimated according to~\mref{eq:39}] the contours are not
very regular circles yet, as they should be for the
independent Gaussians. However, the regular four-peak
structure is clearly visible. As time elapses the structures
with various numbers of peaks appear which represent the
superposition states given by the
formulae~\mref{eq:33}--\mref{eq:38}. The smaller the number
of peaks the better is the separation of the states, and the
more regular is the QPD. The sequence of the pictures is
obtained for $\tau=T_{S}/8,\, T_{S}/6,\, T_{S}/4,\,
T_{S}/3,\,3T_{S}/8$ and $T_{S}/2$, respectively. The
identification of the QPD structures with the corresponding
superposition states is quite obvious. In figure 3 the same
sequence of the QPD structures is shown for the normally
ordered version of the anharmonic oscillator. The differences
between the two versions are quite evident. It is seen that
the period for the normally ordered version is really one
half of the period for the `squared' version. We have chosen
the same sequence of $\tau$ for both versions to visualise
the differences. The structures obtained for the `squared'
version after half the period, which is $T_{S}/2=2\pi$,
become rotations of the initial structures by the angle of
$\pi$, while for the normally ordered version after the same
time the initial structures are recovered. In figure 4 the
contours of the QPD are presented for $\alpha_{0}=4.0$ and
the normally ordered version. Again the structure with the
maximum number of well-separated states, which in this case
is $N_{\max}=8$, shows some irregularity, but as the number of
peaks decreases the structures become more and more regular.
The pictures have been obtained for $\tau=T_{N}/8,\,
T_{N}/6,\, T_{N}/5$ and $T_{N}/3$, respectively. We have
chosen here, as examples, two structures with even number of
peaks and two structures with odd number of peaks, only.
Figures 2-4 are on the same scale, which shows that the
radius of the circle around which peaks are located is in
figure 4 twice as large as that in figures 2 and 3, whereas
the radii of the individual Gaussian bells are the same. One
should also remember that the coefficients of a superposition
with $N$ peaks scale as $|c_{k}|=1/\sqrt{N}$. This means that
the heights of the peaks are N times lower from the single
coherent-state Gaussian. If the initial number of photons
$|\alpha_{0}|^{2}$ becomes large, $|\alpha_{0}|$ is large,
and the maximum number of well-separated states $N_{\max}$ is
also large, but the amplitudes $c_{k}$ of these states become
smaller.

\section{Conclusions}

In this paper we have analysed the process of the generation
of discrete superpositions of coherent states in the course
of the evolution of the anharmonic oscillator. Two versions
of an anharmonic oscillator that are used in the literature
have been compared from the point of view of forming the
superposition states. It has been shown that under an
appropriate choice of the evolution time as a fraction of the
period, the symmetry inherent in the system permits a
considerable simplification of the problem of finding the
coefficients of the superpositions. The number of equations
that must be solved is drastically reduced by the symmetry.
Some examples of the superposition states have been obtained
analytically for superpositions of up to four components.
General rules governing the formation of superpositions with
a definite number of states have been given. The process of
the formation of superposition states has been illustrated
graphically by showing pictures of the corresponding
quasiprobability distributions. The regular structure of the
QPD that is obtained when the time (or length) becomes a
fraction of the period can be easily interpreted as
representing the superposition state that occurs for this
time. The maximum number of well-separated states for given
$|\alpha_{0}|$ has been estimated. This number becomes large
when $|\alpha_{0}|$ becomes large, and regular structures of
the quasiprobability distribution with a large number of
Gaussian peaks can be obtained. Some of these structures have
been shown in the figures. Structures with both even and odd
numbers of peaks are possible. Our results shed some new
light on the problem of generating discrete superpositions of
coherent states and make a contribution to the discussion of
the possibility of generating macroscopically-distinguishable
quantum states~\cite{4,5} as well as to the problem of the
phase space interferences~\cite{16,17}.

\section*{Acknowledgement}

This work was supported by the Polish Research Programme CPBP
01.07.


\begin{thebibliography}{17}

\bibitem{1} Titulaer U and Glauber R J 1965 Phys. Rev. \textbf{145}
1041

\bibitem{2} Stoler D 1971 Phys. Rev. D. \textbf{4} 2309

\bibitem{3} Bia\l{}ynicka-Birula Z 1968 Phys. Rev. \textbf{173} 1207

\bibitem{4} Yurke B and Stoler D 1986 Phys. Rev. Lett. \textbf{57}
13

\bibitem{5} Tombesi P and Mecozzi A 1987 J. Opt. Soc. Am. B \textbf{4}
1700

\bibitem{6} Tana\'s R 1984 {\em Coherence and Quantum Optics V} eds L Mandel
and E Wolf (New York: Plenum) p 645

\bibitem{7} Tana\'s R and Kielich S 1983 Opt. Commun. \textbf{45} 351;
Optica Acta \textbf{31} (1984) 81

\bibitem{8} Milburn G J 1986 Phys. Rev. A \textbf{33} 674

\bibitem{9} Milburn G J and Holmes C A 1986 Phys. Rev. Lett. \textbf{56}
2237

\bibitem{10} Kitagawa and Yamamoto Y 1986 Phys. Rev. A \textbf{34}
3974

\bibitem{11} Pe\v{r}inov\'a V and Luk\v{s} A 1988 J. Mod. Optics \textbf{35}
1513

\bibitem{12} Kennedy T A B and Drummond P 1988 Phys. Rev. A \textbf{38}
1319

\bibitem{13} Sanders B C 1989 Phys. Rev. A \textbf{39} 4284

\bibitem{14} Vourdas A and Bishop R F 1989 Phys. Rev. A \textbf{39}
214

\bibitem{15} Tana\'s R 1989 Phys. Lett. \textbf{141}A 217

\bibitem{16} Schleich W and Wheeler J A 1987 J. Opt. Soc. Am. B \textbf{4}
1715

\bibitem{17} Schleich W, Walls D F and Wheeler J A 1988 Phys. Rev.
A \textbf{38} 1177

\end{thebibliography}
\end{document}